\documentstyle[12pt]{article}
\begin{document}
\begin{center}
CHAOS SYNCHRONIZATION BETWEEN JOSEPSON JUNCTIONS COUPLED WITH TIME DELAYS \\
E.M.Shahverdiev\footnote{Correspong author's e-mail:shahverdiev@physics.ab.az} L.H.Hashimova, P.A.Bayramov and R.A.Nuriev\\ 
Institute of Physics,33,H.Javid Avenue,Baku,AZ1143, Azerbaijan\\
~\\
ABSTRACT\\
\end{center}
We investigate chaos synchronization between Josephson junctions coupled uni-directionally with time-delay. We demonstrate the possibility of high quality synchronization with numerical simulations of such systems. The results are of certain importance for obtaining the high power system of Josephson junctions, which is promising for practical applications.\\
Key words:Josephson junctions; chaos synchronization;time delay systems\\
PACS number(s):05.45.Xt, 05.45 + b, 05.45.Gg, 74.50.+r,74.40.+k\\
~\\
\begin{center}
I.INTRODUCTION
\end{center}
\indent Chaos synchronization [1] is of immense fundamental importance in a variety of complex physical, chemical, power, biological, economical and social systems with possible application areas in secure communications, optimization of nonlinear system performance, modeling brain activity and pattern recognition phenomena, avoiding power black-out, obtaining high power radiation sources, etc.\\
\indent Finite signal transmission times, switching speeds and memory effects makes time-delay systems ubiquitous in nature, technology and society [1-2]. Therefore the study of synchronization phenomena in such systems is of considerable practical importance, see, e.g.[1-3] and references there-in.\\ 
\indent Josepshon junctions are essentially nonlinear dynamical systems [4]. Study of chaos and its control in such systems [4-5] could be of huge practical importance from the application point of view. Chaos control in such systems is important for Josephson junction devices like voltage standards, detectors, SQUID (Superconducting Quantum Interference Device), etc. where chaotic instabilities are not desired. Chaotic Josephson junctions can be used for the short-distance secure communications, ranging purposes, etc [6]. In a chaos-based secure communications a message is masked in the broadband chaotic output of the transmitter and synchronization between the transmitter and receiver systems is used to recover a transmitted message [1-3]. As for ranging, as demonstrated in [7] the high bandwidth and rapid decorrelation of the laser pulse train could provide an unambiguous autocorrelation signal for measuring round-trip delay time. Compared with commercial ladar systems which use the time-of-flight measurement chaotic system use enables better accuracy.\\
\indent It is well-known that Josephson junction is a source of radiation with frequencies up to terahertz region [8]. Terahertz waves, which lie in the part of the electromagnetic spectrum between infrared and microwaves, can penetrate materials that block visible light.They have a wide range of possible uses, including chemical analysis, security scanning, remote sensing chemical signatures of explosives in unopened envelopes, non-invasive detection of cancers in the human body, and high-bandwidth telecommunications [9-10]. In connection with the telecommunications application, it is noted that the earth's atmosphere is a strong absorber of terahertz radiation in specific water vapor absorption bands, so the range of terahertz radiation is limited enough to affect its usefulness in long-distance communications. However, at distances of ~10 meters the band may still allow many useful applications in imaging and construction of high bandwidth wireless networking systems, especially indoor systems. In addition potential uses exist in high-altitude telecommunications, above altitudes where water vapor causes signal absorption: aircraft to satellite, or satellite to satellite.\\
However, the radiation from a single Josephson junction is very weak. Synchronization of the arrays of Josephson junctions is a natural way to increase the radiation power such sources. Upon achieving synchronization the radiation power will be proportional to the number of Josephson junctions squared. Recently it was established that certain highly anisotropic cuprate superconductors 
naturally realise a stack of strongly electromagnetically coupled intrinsic Josephson junctions [8]. From the fundamental point of view synchronization of huge number of complex systems could allow for higher efficiency of computational handling of such systems at least along the synchronization manifolds [1-2]. \\
\indent In this paper we investigate chaos synchronization between three resistively-capacitively(RC)-shunted Josephson junctions coupled uni-directionally with time delays. We demonstrate the possibility of high quality synchronization with numerical simulations of such coupled systems. These results  could be of certain importance for obtaining high power system of Josepson junctions, which is promising for practical applications.\\

\begin{center}
II.CHAOS SYNCHRONIZATION BETWEEN UNIDIRECTIONALLY COUPLED RC-SHUNTED JOSEPHSON JUNCTIONS
\end{center}
\indent Consider synchronization between the following RC-shunted Josephson junctions [5,11] (Fig.1 and Fig.2) written in the dimensionless form:
\begin{equation}
\frac{d^{2}\phi_{1}}{dt^{2}} + \beta_{1} \frac{d\phi_{1}}{dt} + \sin\phi_{1} =i_{dc1} + i_{01}\cos(\Omega_{1}t+\theta_{1})
\end{equation}

\begin{equation}
\frac{d^{2}\phi_{2}}{dt^{2}} + \beta_{2} \frac{d\phi_{2}}{dt} + \sin\phi_{2} =i_{dc2} + i_{02}\cos(\Omega_{2}t +\theta_{2})-\alpha_{s1}(\frac{d\phi_{2}}{dt} -\frac{d\phi_{1}(t-\tau_{2})}{dt})
\end{equation}

\begin{equation}
\frac{d^{2}\phi_{3}}{dt^{2}} + \beta_{3} \frac{d\phi_{3}}{dt} + \sin\phi_{3} =i_{dc3} + i_{03}\cos(\Omega_{3}t +\theta_{3})-\alpha_{s2}(\frac{d\phi_{3}}{dt} -\frac{d\phi_{2}(t-\tau_{3})}{dt})
\end{equation}

Where $\phi_{1},\phi_{2}$,and $\phi_{3}$ are the phase differences of the superconducting order parameter across the 
junctions $1,2$ and $3$, respectively;$\beta$ is called the  damping parameter 
$(\beta R)^2=\hbar(2eI_{c}C)^{-1}$, where $I_{c}, R $ and $C $ are the junctions' critical current, the junction resistance, and capacitance, respectively;$\beta$ is related to McCumber parameter $\beta_{c}$ by $\beta^{2}\beta_{c}=1$;
$\hbar$ is Planck's constant divided by $2\pi$;$e$ is the electronic charge;$i_{dc1,dc2,dc3}$ is the driving the junctions direct current;$i_{01,02,03}\cos(\Omega_{1,2,3} t+\theta_{1,2,3})$ is the driving ac (or rf) current with amplitudes $i_{01,02,03}$,frequencies $\Omega_{1,2,3}$ and phases $\theta_{1,2,3}$;$\tau_{1}$ and $\tau_{2}$ are the coupling delay times between the junctions $1-2$ and $2-3$, respectively;coupling between the junctions $1-2$ and $2-3$ is due to the currents flowing through the coupling resistors $R_{s1}$ (between the junctions $1$ and $2$) and $R_{s2}$(between the junctions $2$ and $3$);$\alpha_{s1,s2}=R_{1,2}\beta_{1,2}R_{s1,s2}^{-1}$ are the coupling strengths between junctions $1$-$2$ and junctions $2$ - $3$. We note that in Eqs.(1-3)direct current and ac current amplitudes are normalized with respect to the critical currents for the relative  Josephson junctions; ac current frequencies $\Omega_{1,2,3}$ are normalized with respect to the Josephson junction plasma frequency $\omega^{2}_{J1,J2,J3}=2eI_{c1,c2,c3}(\hbar C_{1,2,3})^{-1},$ and dimensionless time is normalized to the inverse plasma frequency. \\
\indent It is noted that if the external drive current is purely dc then we have a second order autonomous system, Eq.(1) and therefore chaotic dynamic for Eq. (1) is ruled out. In order to make the dynamics of the Josephson junction (1) chaotic, one can add ac current to the external driving, which makes Eq.(1) a second order non-autonomous system. Then treating  $\Omega_{1} t$ term as a new dynamical variable Eq.(1) (and consequently Eqs.(2-3)) can be rewritten as a system of third-order ordinary differential equations, which can behave chaotically.
\begin{equation}
\frac{d\phi_{1}}{dt}=\psi_{1}
\end{equation}
\begin{equation}
\frac{d\psi_{1}}{dt}= - \beta_{1} \psi_{1} - \sin\phi_{1} + i_{dc1}+ i_{01}\cos \varphi_{1}
\end{equation}
\begin{equation}
\frac{d\varphi_{1}}{dt}=\Omega_{1} 
\end{equation}
\begin{equation}
\frac{d\phi_{2}}{dt}=\psi_{2}
\end{equation}
\begin{equation}
\frac{d\psi_{2}}{dt}= - \beta_{2} \psi_{2} - \sin\phi_{2} + i_{dc2} + i_{02}\cos \varphi_{2}
 -\alpha_{s1}(\psi_{2} -\psi_{1}(t-\tau_{2}))
\end{equation}
\begin{equation}
\frac{d\varphi_{2}}{dt}=\Omega_{2}
\end{equation}
\begin{equation}
\frac{d\phi_{3}}{dt}=\psi_{3}
\end{equation}
\begin{equation}
\frac{d\psi_{3}}{dt}= - \beta_{3} \psi_{3} - \sin\phi_{2} + i_{dc3} + i_{03}\cos \varphi_{3}
 -\alpha_{s2}(\psi_{3} -\psi_{2}(t-\tau_{2}))
\end{equation}
\begin{equation}
\frac{d\varphi_{3}}{dt}=\Omega_{3}
\end{equation}
We underline that the value of phases $\theta_{1,2,3}$  for the ac driving forces can be incorporated into the initial conditions for Eqs.(4-12).\\
\indent We note that Eq.(1) is also used for modeling of a driven nonlinear pendulum, charge density waves with a torque and sinusoidal driving field [12-13]. We also mention that there are some other ways to make the Josephson dynamics chaotic without resorting to the ac driving. One way is to use resistively-capacitively-inductively-shunted (RCL-shunted) Josephson junction [5]. Additionally one can make the Josephson junction dynamics chaotic via time- delays in the system [11].\\
\begin{center}
III.NUMERICAL SIMULATIONS
\end{center}
Now we demonstrate that three unidirectionally time delay coupled Josephson junctions can be synchronized.
We use DDE software in  Matlab 7  for numerical simulations. We study chaos synchronization between Eqs.(4-12) using the correlation coefficient $C$ [14] 
$$C(\Delta t)= \frac{<(x(t) - <x>)(y(t+\Delta t) - <y>)>}{\sqrt{<(x(t) - <x>)^2><(y(t+ \Delta t) - <y>)^2>}}$$
where $x$ and $y$ are the outputs of the interacting systems; the brackets$<.>$
represent the time average; $\Delta t$ is a time shift between the systems outputs: in our case$\Delta t=0.$   This coefficient indicates the quality of synchronization: $C=1$ means perfect synchronization.\\
First we simulate Eqs.(4-12) for the identical set of systems' parameters:$\beta_{1}=\beta_{2}=\beta_{3}=0.25,i_{dc1}=i_{dc2}=i_{dc3}=0.3,i_{01}=i_{02}=i_{03}=0.7,\Omega_{1}= \Omega_{2}= \Omega_{3}=0.6, \theta_{1}=\theta_{2}=\theta_{3}=0, \alpha_{s1}= \alpha_{s2}=0.45, \tau_{1}=\tau_{2}=0.15.$Throught the paper initial conditions are set to be different for 
the junctions.\\
Fig.3(a) demonstrates dynamics of variable $\psi_{1}$;fig.3(b)pictures the error dynamics $\psi_{2}-\psi_{1}$;$C_{12}$  is the  correlation coefficient between $\psi_{2}$ and $\psi_{1}.$ Fig.3(c) represents the dependence between variables $\psi_{3}$ and $\psi_{2}$ with $C_{23}$ being the correlation coefficient between $\psi_{2}$ and $\psi_{3}.$ In fig.3(d) we present the error dynamics between the end junctions $\psi_{3}-\psi_{1}$, while $C_{23}$ indicates the correlation coefficient between $\psi_{1}$ and $\psi_{3}.$ These simulation results underline the high quality synchronization between three identical Josephson junctions with different initial conditions.\\
Next we study the effect of the coupling delay time on the synchronization quality. It is worth noting that synchronization quality is very sensitive to the value of the coupling delay. As demonstrated by the numerical simulations 
with increasing coupling delays between the Josephson junctions correlation coefficients decays rapidly.In fig.4 we present the dependence of the correlation coefficients $C_{12},C_{23}$ and $C_{13}$ on the time delay. It is noted that we only increase the values of the coupling delay times, but keep them equal: $\tau_{1}=\tau_{2}.$ \\
Fig.5 shows the results of numerical simulations of chaos synchronization between 9 coupled Josephson Junctions  with $\tau_{1}=\tau_{2}=\cdots =\tau_{8}=0.001.$ and the other parameters are as in fig.3: error dynamics of $\psi_{9}-\psi_{1}$:$C_{19}$ is the correlation coefficient between $\psi_{1}$ and $\psi_{9}.$
\indent Finally we briefly dwell on the role of noise in the synchronization. Noise can both create and deteriorate the synchronization quality between the systems. It all depends on the intensity of noise and parameter mismatches between the systems [15]. Detailed study of the influence of noise on the synchronization quality will be reported in the future.\\ 
\begin{center}
IV.CONCLUSIONS
\end{center}
\indent We have numerically simulated chaos synchronization between three uni-directionally time-delay coupled Josephson junctions. We have demonstrated the possibility of high quality synchronization between such systems.  The results are important for obtaining high power system of Josephson junctions, which is promising for practical applications.\\
\newpage
\begin{center}
FIGURE CAPTIONS
\end{center}
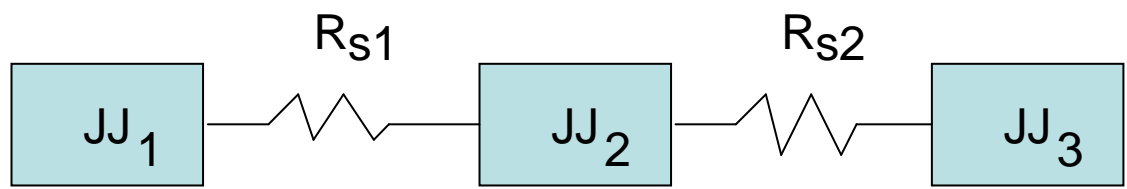
FIG.1.Schematic view of the three unidirectionally coupled Josephson Junctions $JJ_{1}$, $JJ_{2}$ and $JJ_{3}.$ Coupling is realised via the resistors $R_{s1}$ and $R_{s2}.$\\
~\\
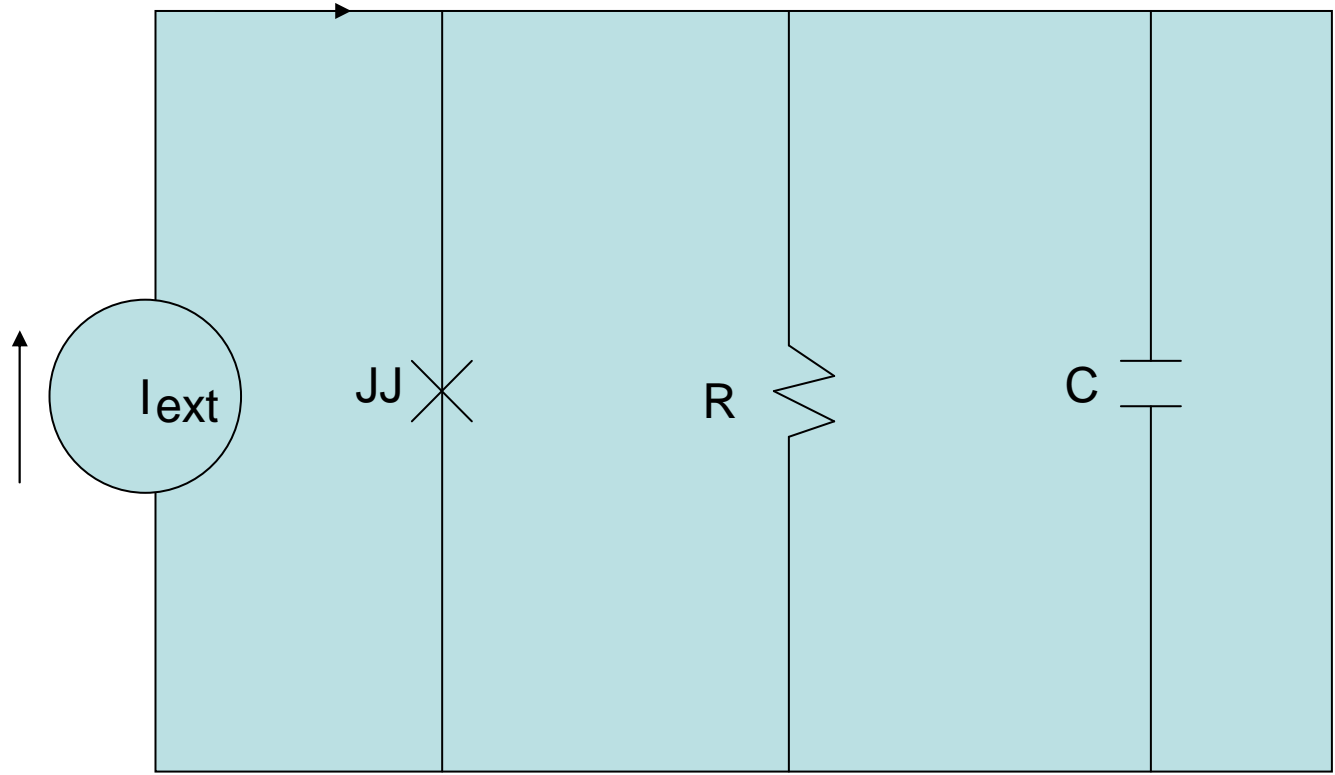
FIG.2.Schematic set-up of the RC-shunted Josephson Junction subject to the external driving $I_{ext}.$\\
~\\
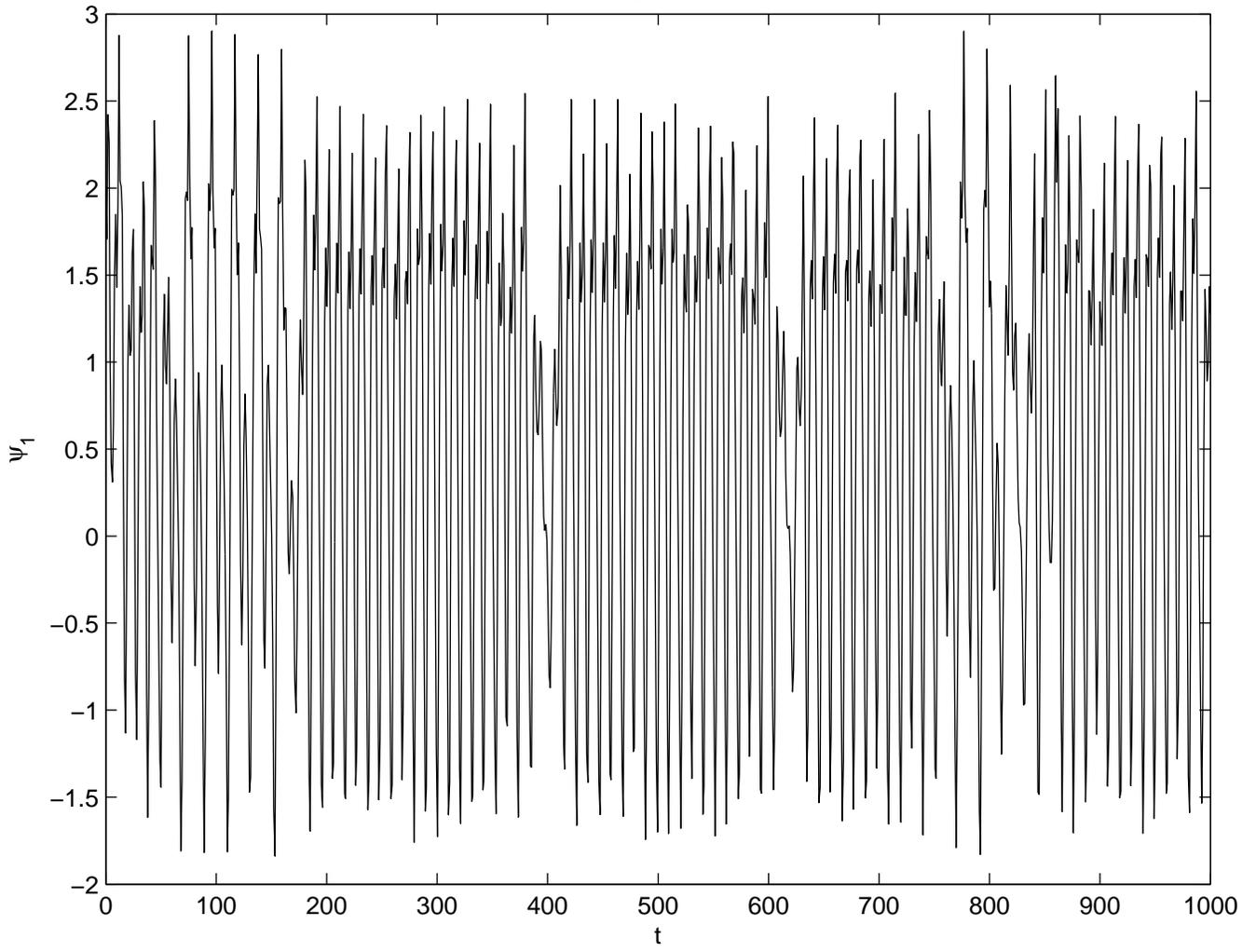
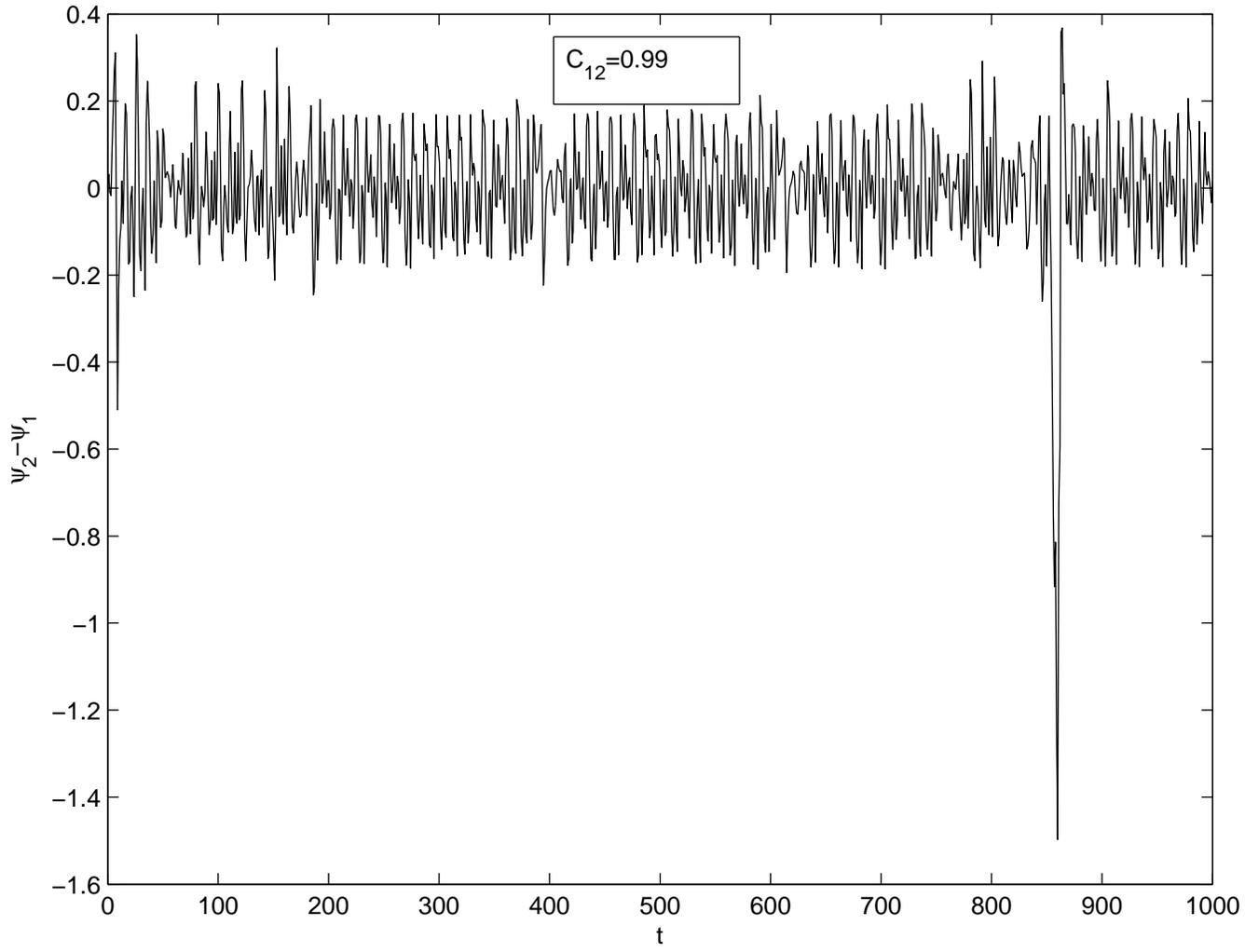
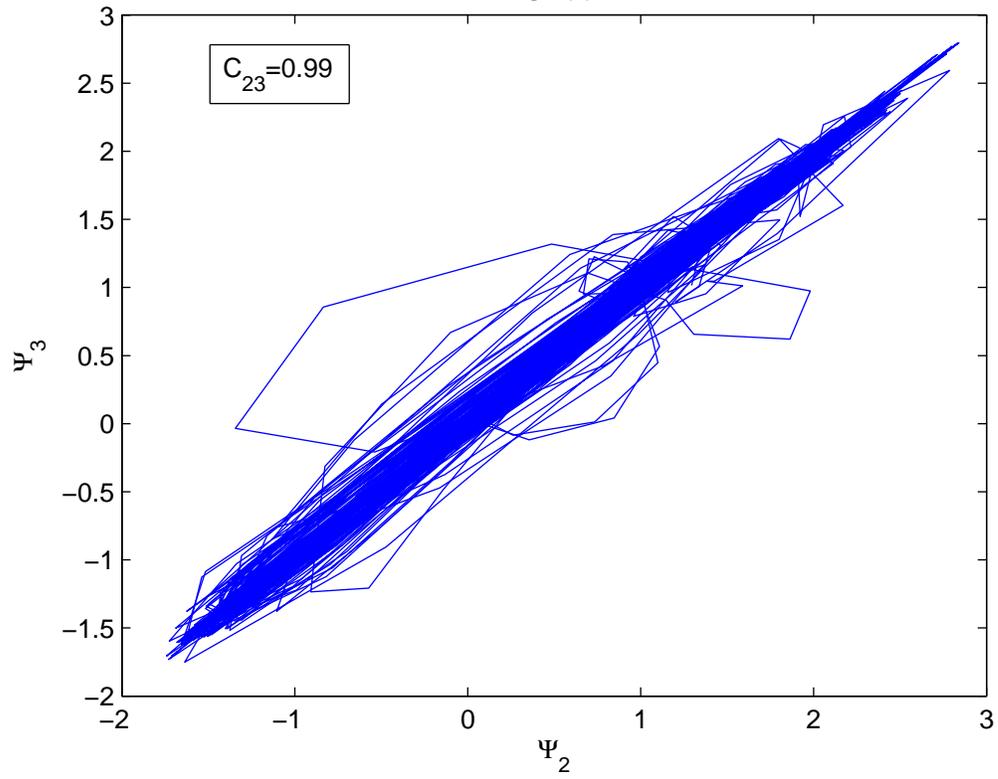
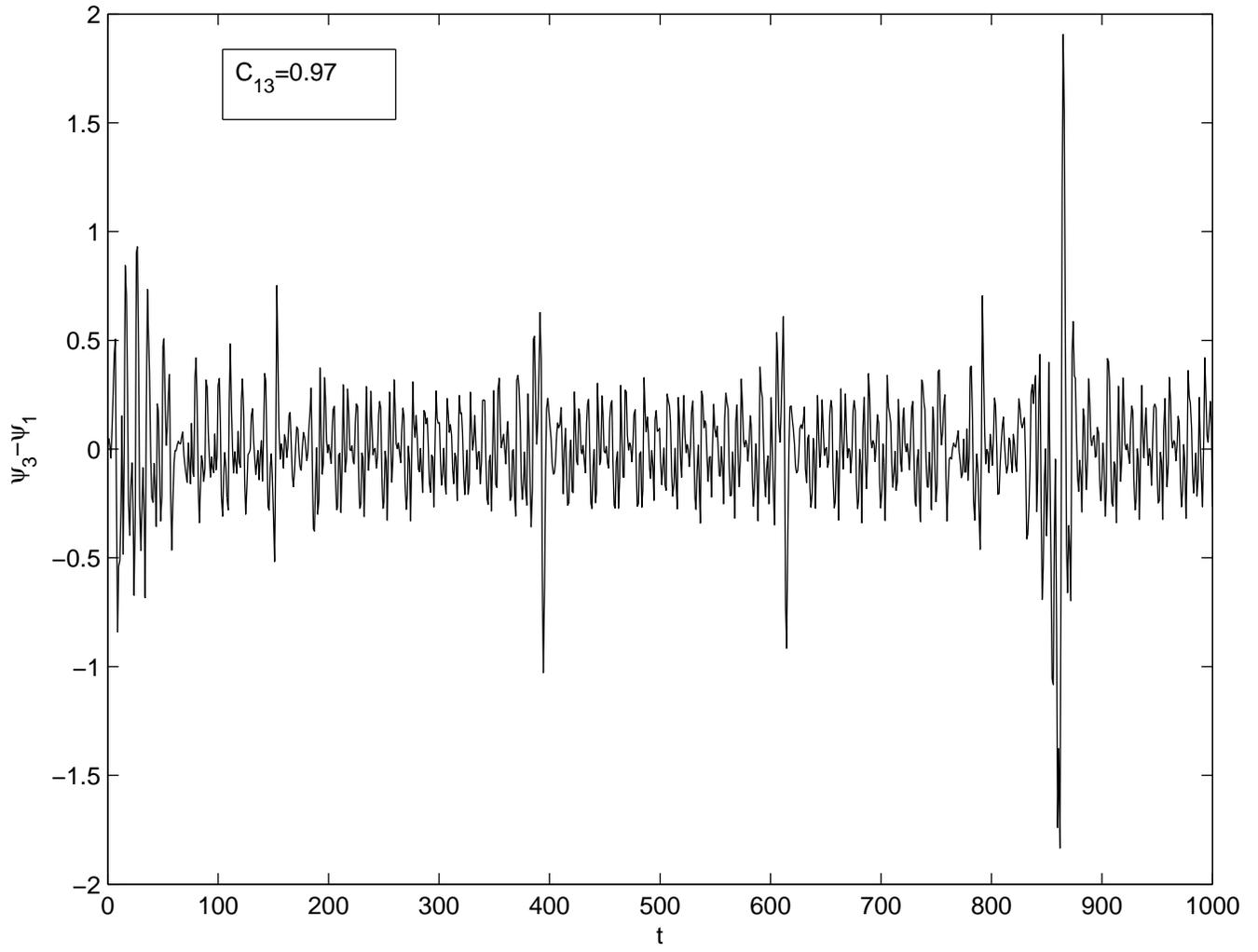
FIG.3.Numerical simulation of Eqs.(4-12):(a)time series of $\psi_{1}$;(b)error $\psi_{2}-\psi_{1}$ dynamics; $C_{12}$ is the correlation coefficient between the Josephson Junctions $JJ_{1}$ and $JJ_{2}$;(c)dependence between variables $\psi_{3}$ and $\psi_{2}$: $C_{23}$ is the correlation coefficient between $\psi_{2}$ and $\psi_{3}$;(d)error dynamics of $\psi_{3}-\psi_{1}$:$C_{13}$ is the correlation coefficient between $\psi_{3}$ and $\psi_{1}.$ The parameters are:$\beta_{1}=\beta_{2}=\beta_{3}=0.25,i_{dc1}=i_{dc2}=i_{dc3}=0.3,i_{01}=i_{02}=i_{03}=0.7,\Omega_{1}= \Omega_{2}= \Omega_{3}=0.6, \theta_{1}=\theta_{2}=\theta_{3}=0, \alpha_{s1}= \alpha_{s2}=0.45, \tau_{1}=\tau_{2}=0.2.$ Dimensionless units.\\
~\\
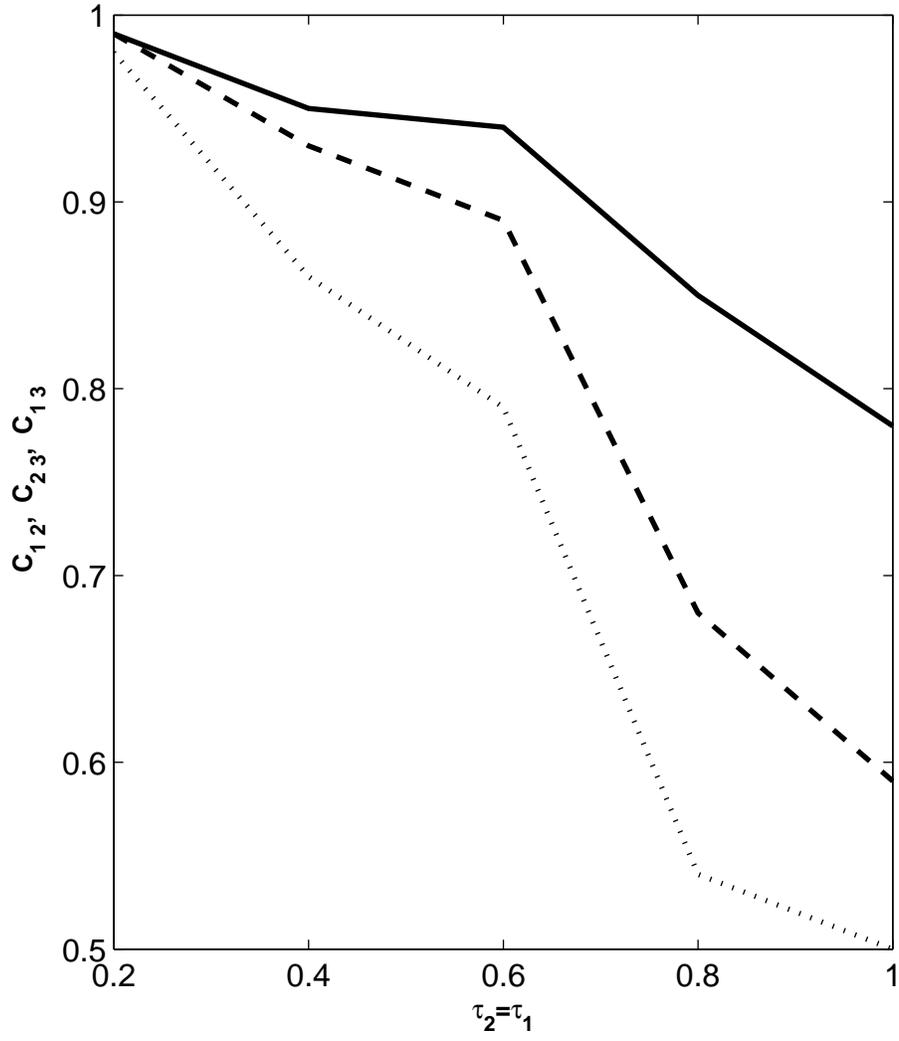
FIG.4.Numerical simulation of Eqs.(4-12):Dependence of the correlation coefficients $C_{12}$ (solid line),$C_{23}$(dashed line) and $C_{13}$ (dotted line) on the coupling delay time between the junctions $\tau_{1}=\tau_{2}.$ Other parameters are as in figure 3. Dimensionless units. \\
~\\
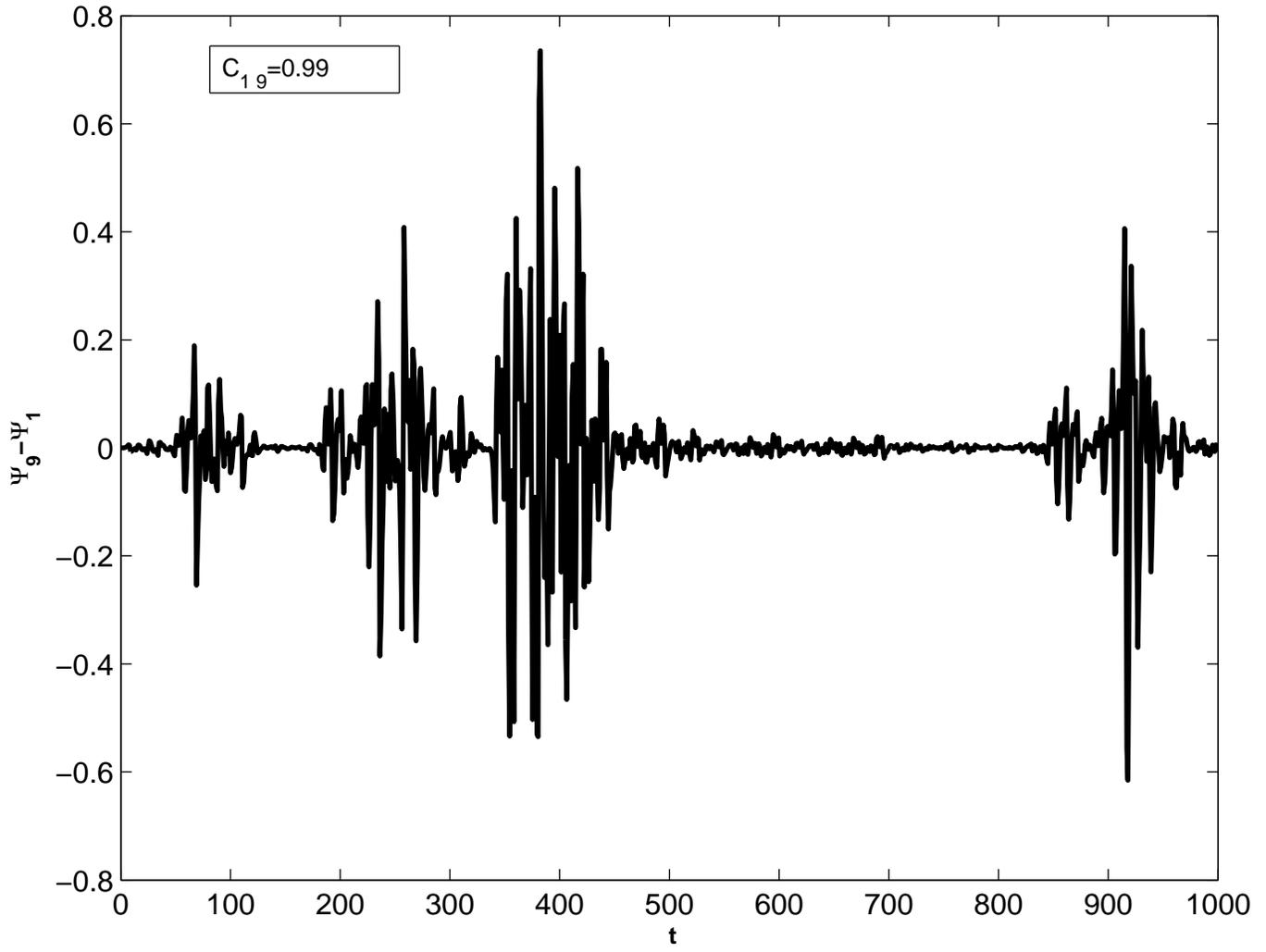
FIG.5.Numerical simulation of chaos synchronization between 9 coupled Josephson Junctions  with $\tau_{1}=\tau_{2}=\cdots =\tau_{8}=0.001.$ and the other parameters are as in fig.3: error dynamics of $\psi_{9}-\psi_{1}$:$C_{19}$ is the correlation coefficient between $\psi_{1}$ and $\psi_{9}.$ Dimensionless units.\\

\end{document}